\documentclass[aps,pra,twocolumn,superscriptaddress,floatfix]{revtex4}
\usepackage{graphicx,dcolumn,bm,hyperref,amsmath,amssymb,xspace,epsfig,float,array,multirow,amsfonts}
\usepackage{xcolor}
\begin{document}
\title{\bf Scaling properties of the Tan's contact: embedding pairs and correlation effect in the Tonks-Girardeau limit}

\author{F. T. Sant'Ana}
\affiliation{Universit\'e C\^ote d'Azur, CNRS, Institut de Physique de Nice, France}
\affiliation{S\~ao Carlos Institute of Physics, University of S\~ao Paulo, 13566-590, S\~ao Carlos, SP, Brazil}
\author{F. H\'ebert}
\affiliation{Universit\'e C\^ote d'Azur, CNRS, Institut de Physique de Nice, France}
\author{V. G. Rousseau}
\affiliation{5933 Laurel St, New Orleans, LA70115, USA}
\author{M. Albert}
\affiliation{Universit\'e C\^ote d'Azur, CNRS, Institut de Physique de Nice, France}
\author{P. Vignolo}
\affiliation{Universit\'e C\^ote d'Azur, CNRS, Institut de Physique de Nice, France}
\date{\today}

\begin{abstract}
  We study the Tan's contact of a one dimensional quantum gas of $N$ repulsive identical bosons
  confined in a harmonic trap at finite temperature. This canonical
  ensemble framework corresponds to the experimental conditions, the number
  of particles being fixed for each experimental sequence. We show that, in
  the strongly interacting regime, the contact rescaled by the contact at the
  Tonks-Girardeau limit is an universal function of two parameters, the rescaled
  interaction strength and temperature. This means that all pair and
  correlation effects in the Tan's contact are embedded in
  the Tan's contact in the Tonks-Girardeau limit.

 % This claim is valid for any number
 % of particles, from two to infinity, and any temperature,
 % from zero to infinity.
 
\end{abstract}
\maketitle
\section{Introduction}

Many-body quantum physics is a cornerstone of modern physics and a key
to understand future technologies such as high $T_c$ superconductivity or
quantum computing. However, an accurate description of
strongly correlated quantum systems, for an arbitrary number of particles,
is often a dare without a simple solution. Apart from the very specific
family of integrable systems \cite{Lieb1963,McGuire1964,Yang1969,LiebLiniger,Yang67,Gaudin1967,Sutherland68,LaiYang,McGuire,Luther1974,Fuchs2004} where all observables can, in principle, be predicted theoretically,
our knowledge is in general limited to simple situations like two particles
\cite{Busch98,Aharony2000,Abad2005}, solutions that hold in the
thermodynamic limit \cite{Olshanii03,Yao2018}, low energy
physics \cite{Giamarchi_book}, or mean-field descriptions for many-body
systems \cite{Gross1961,Pitaevskii1961}. It is therefore quite delicate
to extract general informations such as the scaling of physical observables
with respect to the number of particles for generic situations.

For the case of quantum particles with point-like interactions,
short-range correlations are embedded in the Tan's contact $C_N$
\cite{Tan2008a,Tan2008b,Tan2008c}.
This quantity, that is proportional to
the probability that two particles approach each other infinitely close,
determines the asymptotic behaviour of the momentum distribution $n(k)$,
$C_N=\lim_{k\rightarrow\infty}k^4n(k)$, $k$ being the momentum divided by $\hbar$.
This observable can be measured via
time-of-flight techniques \cite{Stewart2010,Sagi2012,Chang2016},
with radio-frequency spectroscopy \cite{Wild2012,Yan2019},
Bragg spectroscopy \cite{Hoinka2013},
by measuring the energy
variation as a function of the interaction strength \cite{Sagi2012}, or by
looking at three-body losses in quantum mixtures \cite{Laurent2017}.
This central quantity is a function of the interaction energy, density-density
correlations function, the trapping configuration, the temperature as well as
the magnetization \cite{Decamp2016b,Decamp2017}, and thus depends in a non
trivial way on the nature and the number $N$ of particles. Therefore, even in one
dimension, the behaviour of $C_N$ is not completely clarified,
especially in trapped systems, despite many
theoretical investigations \cite{Minguzzi02,Lewenstein-Massignan,Matveeva2016,
  Patu2016,Decamp2016b}. For one-dimensional (1D) bosons (and/or fermions)
trapped in a harmonic potential of frequency $\omega$,
it has been shown that, in the thermodynamic
limit, at zero temperature, the contact rescaled by $N^{5/2}$ is
a universal function of one scaling parameter: $z=a_{ho}/(|a_{1D}|\sqrt{N})$
\cite{Olshanii03,Matveeva2016}. This holds also at finite temperature,
in the grand-canonical ensemble: the contact rescaled by $N^{5/2}$ is
a universal function of two scaling parameters, $z$
and $\xi_T=|a_{1D}|/\lambda_{DB}$, or equivalently $z$ and $\tau=T/T_F$
\cite{xu2015,Yao2018}, $a_{1D}$ being the 1D scattering length, $a_{ho}=\sqrt{\hbar/(m\omega)}$ the
harmonic oscillator length, $m$ being the mass,
$\lambda_{DB}=\sqrt{2\pi\hbar^2/mk_BT}$
the De Broglie thermal wavelength, $T_F=N\hbar\omega/k_B$
the Fermi temperature, and $k_B$ the Boltzmann constant.
However, for systems with small number of particles,
the $N^{5/2}$-scaling fails. In the  zero-temperature limit \cite{Rizzi2018},
it is possible to change the paradigm and to introduce a different scaling form
that holds from $N=2$ to infinity. At finite temperature, in the
grand-canonical ensemble, the $N^{5/2}$-scaling holds for $N>10$ \cite{Yao2018}.
However, corrections at small number of particles have, to our knowledge,
not yet been studied in 1D, and the important question of the relevance of
the statistical ensemble has not been addressed. The latter is indeed a
crucial point since ultracold atom experiments are canonical or, more often,
an average over canonical ensembles, but not grand canonical and scaling
properties are obviously strongly affected by the statistical distribution
of particles numbers. In fact, in ultracold experiments, in each experimental
sequence, $N$ atoms are charged
in a three-dimensional trap. Then the atoms are separated in several
light wires created by the interference of two propagating
laser beams \cite{Moritz2003}.
The atomic gas in the wires can be considered as one-dimensional,
if the interaction and thermal energies are lower
than the energy scale of the radial confinement $\hbar\omega_\perp$,
$\omega_\perp$ being the radial
harmonic oscillator frequency \cite{Pagano2014}.
Otherwise, atoms can be directly trapped in a single 1D tube with a
strong radial confinement \cite{Salces2018}.
In both cases, the relation between the 1D scattering length $a_{1D}$
and the 3D one $a_{3D}$ is given by $a_{1D}=-a_\perp^2/a_{3D}$,
where $a_\perp=\sqrt{\hbar/(m\omega_\perp)}$ \cite{Olsh98}.

In this paper we study the canonical Tan's contact
for a small number of harmonically trapped Lieb-Liniger bosons.

We show that, in the strongly interacting regime, the contact for $N$ bosons
at temperature $T$  and with repulsive
interaction, divided by the
contact for the same number of bosons and temperature but
in the regime of infinite repulsions,
is a $N$-independent function of $z$ and $\tau$.
Namely, all the
non-trivial particle-number dependence is embedded in
the contact in the infinite interaction limit, even at finite temperature,
which is the main result of this work.
The regime of infinite repulsions in one-dimension corresponds
to the so-called Tonks-Girardeau limit. In this regime,
the infinite repulsions,
due to the low-dimensionality,
play the role of a sort of Pauli principle
so that bosons ``behave'' as non-interacting fermions.
Another result is that we provide an analytical expression
for the $N$-dependence of the canonical contact in the Tonks-Girardeau limit.
Our formula is a conjecture that works extremely well over the whole
temperature range.
The consequence of these two results is that we can 
explicitly express the canonical contact for $N$ harmonically
trapped Lieb-Liniger bosons in the intermediate and
strong-interaction regime ($z>1$),
for any value of $N$ and any temperature $T$.

The paper is organized as follows.
In Sec. \ref{sec-can} we introduce the physical system
and define the canonical Tan's contact.
This observable is then evaluated exactly in two special situations: for two identical bosons at any interaction strength and any temperature and
for $N$ identical bosons in the Tonks-Girardeau limit (infinite coupling). %In this latter case we provide an explicit analytical expression. 
In the general situation, namely for intermediate interaction strength and for $N>2$, we calculate the Tan's contact by means of Quantum Monte Carlo (QMC) simulations.
The scaling properties of the canonical contact are then analyzed in Sec.~\ref{sec-scaling}. After reminding the results previously obtained,
in the strongly-interacting limit, at zero temperature \cite{Rizzi2018},
we analyze the large temperature scaling of the contact in the same limit.
By comparing these two limits, we propose an explicit form of the contact scaling function holding
in the strongly interacting limit and at any temperature which makes
our numerical data overlap for different number of atoms $N$ with only a few percent discrepancy. In Sec. \ref{sec-comp} we compare the canonical contact with the grand-canonical one. At large temperature the canonical and grand-canonical contacts are both proportional to the two-bosons contact. This does not hold at smaller temperatures. Finally, our concluding remarks are given in Sec. \ref{sec-concl}.

%\begin{equation}
%C_N^{gc}(g,T)/N^{5/2}=f(z,\xi_T)=\tilde f(z,\tau).
%\end{equation}
\section{Canonical Tan's contact}
\label{sec-can}
We consider a gas of $N$ identical interacting bosons of mass $m$ trapped in a 1D harmonic
confinement. This system is described by the Hamiltonian
\begin{equation}
H=\sum_{i=1}^{N}\left(-\frac{\hbar^{2}}{2m}\frac{\partial^2}{\partial x_{i}^{2}}
	+\frac{1}{2}m\omega^{2}x_{i}^{2}\right)
	+g\sum_{i<j}\delta\left(x_i-x_j\right), \label{hamiltonian}
\end{equation}
where the repulsive interaction strength $g$ depends on the 1D scattering length
as $g=-2\hbar^2/ma_{1D}$, if $a_\perp\gg a_{3D}$ \cite{Olsh98}.
At finite temperature $T$, in the canonical ensemble,
the contact for $N$ bosons, $C_N^c(g,T)$, can be deduced from
the free energy $F$ by exploiting the Tan's sweep relation \cite{Tan2008a}
\begin{equation}
  \begin{split}
  C_N^c(g,T)=&-\dfrac{m^2}{\pi\hbar^4}\dfrac{\partial F}{\partial g^{-1}} \\
  =&-\dfrac{m^2}{\pi\hbar^4}\dfrac{\sum_ie^{-\beta E_i}
    \partial E_i/\partial g^{-1}}{\sum_ie^{-\beta E_i}},\\
  \label{def-can}
\end{split}
\end{equation}
where $E_i$ is the $i$-th eigenenergy of the $N$-boson system and $\beta=(k_BT)^{-1}$.
$C_N^c(g,T)$ can be exactly evaluated for $N=2$ at any value of the interaction strength $g$ and any temperature $T$,
and in the Tonks-Girardeau limit $g\rightarrow\infty$ for any $N$ and $T$.

Let us underline that, analogously to the zero-temperature case,
the contact can also be calculated from the average interaction energy
that can be obtained by the free energy
from the Hellmann-Feynman theorem
$\langle H_{\rm int}\rangle=g\partial F/\partial g$ \cite{Valiente2012}.
It follows \cite{Tan2008b}
\begin{equation}
C_N^c(g,T) = \frac{gm^2}{\pi \hbar^4} \langle H_{\rm int}\rangle.\label{Hint}
\end{equation}

\subsection{The two bosons system}
\begin{figure}
\centerline{\includegraphics[width=1.\linewidth]{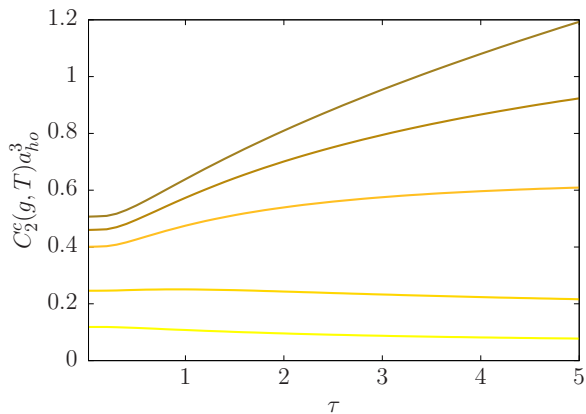}}
  \caption{\label{felipe} Canonical Tan's contact $C^c_2(g,T)$ as a function of $\tau=T/T_F$ [Eq.(\ref{two-bos})] for different values of the interaction strength $z=a_{ho}/(|a_{1D}|\sqrt{N})$. From bottom to top: $z=0.5$, 1, 2.5, 5, and 1000. The curve for $z=1000$ is indiscernable from the contact evaluated in the Tonks limit by means of Eq. (\ref{two-bos-inf}).}
\end{figure}

For the two bosons system, the energy spectrum can be
calculated analytically. In this case $E_i=E_{cm,\ell}+E_{r,j}$,
$E_{cm,\ell}$ being the
centre of mass energy with quantum number $\ell$
and $E_{r,j}=\hbar\omega(1/2+\nu_j)$ the relative energy, with quantum number
$j$ [$i=(\ell,j)$], that depends on the
interaction strength via the implicit relation \cite{Busch98}
\begin{equation}\label{eq.gamma}
  f(\nu)=\frac{\Gamma\left(-\frac{\nu}{2}\right)}
	{\Gamma\left(-\frac{\nu}{2}+\frac{1}{2}\right)} = -\sqrt{2}\frac{|a_{1D}|}{a_{ho}},
\end{equation}
where $\Gamma(x)$ is the gamma function \cite{Gradshteyn}.
$E_{cm,\ell}$, differently from the relative energy $E_{r,j}$, is completely independent on interatomic
interactions as stated by the Kohn's theorem \cite{Pitaevskii2016} and then does not contribute to the
contact calculation.
By applying Eq. (\ref{def-can}), the two bosons contact then takes the form
\begin{equation}
  \begin{split}
	  C_2^c(g,T) =&\frac{\sqrt{8} z^2}{\pi a_{ho}^3} Z_r^{-1} \sum_{j}e^{-\beta \hbar\omega\nu_j}\frac{\partial\nu_j}{\partial z} \\
	  =&\frac{\sqrt{32}}{\pi a_{ho}^{3}} Z_r^{-1} \sum_{j}e^{-\beta \hbar\omega \nu_j} 
	  \frac{\Gamma\left(-\frac{\nu_j}{2}+\frac{1}{2}\right)}{\Gamma\left(-\frac{\nu_j}{2}\right)}\\
	  &\times \left[\psi\left(-\frac{\nu_j}{2}+\frac{1}{2}\right)-\psi \left(-\frac{\nu_j}{2}\right)\right]^{-1},
  \end{split}
  \label{two-bos}
\end{equation}
where $Z_r=\sum_{j}e^{-\beta\hbar\omega\nu_j}$ is the canonical relative motion partition function
and $\psi(x)=\Gamma^\prime(x)/\Gamma(x)$ is the digamma function \cite{Gradshteyn}.
In the Tonks-Girardeau limit $\nu_j=2j-1$ ($j\ge 1$) and both $\Gamma\left(-\frac{\nu_j}{2}+\frac{1}{2}\right)=\Gamma(-j+1)$
and $\psi\left(-\frac{\nu_j}{2}+\frac{1}{2}\right)=\psi(-j+1)$
diverge for $j\ge1$.
With some algebra, it can been shown that
\begin{equation}
  C_2^c(\infty,T)=\frac{\sqrt{32}}{\pi^{3/2} a_{ho}^{3}} Z_r^{-1} \sum_{j}e^{-\beta \hbar\omega (2j-1)} \dfrac{(2j-1)!!}{2^j(j-1)!}
  \label{two-bos-inf}
\end{equation}  
Remark that Eq. (\ref{two-bos-inf}) gives
the known limit $C_2^c(\infty,0)=(2/\pi)^{3/2}a_{ho}^{-3}$ \cite{Rizzi2018}.
The canonical two-bosons contact obtained by Eq. (\ref{two-bos})
is shown in Fig. \ref{felipe}.
We have verified that the curve for $z=1000$ is essentially
indiscernable from the contact evaluated in the Tonks limit by
means of Eq. (\ref{two-bos-inf}).

\medskip

\subsection{The Tonks-Girardeau limit}
In the Tonks-Girardeau limit, where fermionization occurs, the interaction strength
$g$ is ininite, namely the 1D scattering length
$a_{1D}$ is zero and therefore, this length-scale disappears by making the
problem more universal.
Thus the contact, in this regime, does not depend on the interactions
and can be written
as a function of the corresponding fermionic two-body density matrix
$\rho_{2F} (x_1,x_2;x_1',x_2')$ \cite{Fang09}.
More precisely, it can be shown that
\begin{equation}
  C_N^c(\infty,T)=\dfrac{2}{\pi}\int_{-\infty}^{+\infty}{\rm d}x F(x)
  \label{miaeq}
\end{equation}  
where we have defined
\begin{equation}
F(x)=\lim_{x',x''\rightarrow x} \dfrac{\rho_{2F} (x',x;x'',x)}{|x-x'||x-x''|}.
\end{equation}
By explicitly expressing $\rho_{2F}$ in the canonical ensemble,
as a function of the single-particle orbitals $u_i(x)$, we get
%\begin{widetext}
\begin{equation}
  \begin{split}
 & F(x)=Z^{-1}\sum_{\substack{i_1=0,\infty, i_2=i_1+1,\infty\\
        \dots
        i_{N_F}=i_{N_F-1}+1,\infty}}e^{-\beta\hbar\omega \sum_{j=1,N_F}(i_j+\frac{1}{2})}\\
 &   \sum_{\langle j,k\rangle} \left([u_{i_j}(x)\partial_x u_{i_k}(x)]^2
    -2 u_{i_j}(x)\partial_x u_{i_k}(x)u_{i_k}(x)\partial_x u_{i_j}(x)\right)\\
    \end{split}
  \label{miaancora}
\end{equation}
%\end{widetext}
with
\begin{equation}
Z=\sum_{\substack{i_1=0,\infty,i_2=i_1+1,\infty\\
        \dots
        i_{N_F}=i_{N_F-1}+1,\infty}}e^{-\beta\hbar\omega\sum_{j=1,N_F}(i_j+\frac{1}{2})}.
\end{equation}
    \begin{figure}
\begin{center}
  \includegraphics[width=1.\linewidth]{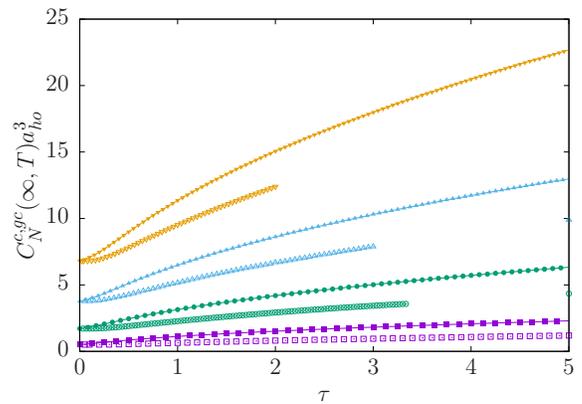}
  \caption{\label{Tonks-c-gc} Canonical (empty symbols) [Eq.~(\ref{miaeq})]
    and grand-canonical contact (full symbols) \cite{vignolo2013}
    as a function of $\tau$
    for $N=2$ (violet squares),
    $N=3$ (green circles), $N=4$ (light-blue up-triangles), and $N=5$ (orange down-triangles) Tonks-Girardeau bosons. The grand-canonical case will be discussed in Sec. \ref{sec-comp}.}
    \end{center}
      \end{figure}
    The canonical contact $C^c_N(\infty,T)$,
    as obtained by Eqs. (\ref{miaeq})
    and (\ref{miaancora}), is shown in Fig. \ref{Tonks-c-gc} (empty symbols)
    for $N=2$ to 5. The data are compared with grand-canonical
    ones \cite{vignolo2013} (full symbols) that will be discussed below (Sec. \ref{sec-comp}).
    Remark that the computation of the
    contact is more demanding in the canonical case than in the grand-canonical one, because of 
    several sums in (\ref{miaancora}) that simplify in the grand-canonical case.

\subsection{The finite interaction strength regime}

In the finite interaction strength scenario for $N>2$,
we rely on quantum Monte Carlo simulations to obtain
exact results. 
Starting from Eq. (\ref{hamiltonian}), we discretize the Hamiltonian using a finite difference
method and rewrite it using second quantization, ending with the following bosonic Hubbard
Hamiltonian
\begin{eqnarray}
H &=& -t \sum_j \left( b^\dagger_{j}b_{j+1}  -2 n_j + b^\dagger_j b_{j-1}\right)  \nonumber\\
&& +w \sum_j j^2 n_j  + U \sum_j n_j(n_j-1)/2. \label{hubbard}
\end{eqnarray}
The discrete positions of the bosons are given by $x= j\Delta a_{ho}$ where $\Delta$ is
a small dimensionless parameter.
We typically used $\Delta=0.1$ and checked on some simulations
that the systematic errors induced by this discretization were smaller than the stochastic 
errors due to the Monte Carlo calculations.
The operators $b^\dagger_{j}$ and $b_j$ create or destroy bosons on site $j$. $n_j = b^\dagger_j b_j$
is the bosonic number operator on site $j$. The parameters are given by
\begin{equation}
t = \frac{\hbar \omega}{2\Delta^2},\  w = \frac{\hbar \omega \Delta^2}{2},\ U= \frac{g}{\Delta a_{ho}}.
\end{equation}

The Hubbard model is simulated using the stochastic Green function algorithm
\cite{rousseau08,rousseau08b} that allows the calculation of many physical quantities for finite systems at finite
temperature. The algorithm works in both canonical and grand-canonical ensembles, although it is generally
more efficient in the former case. Grand canonical simulations require the sampling of a larger
space containing different numbers of particles, which increases a lot the correlation time of the data, as
the sampling of different $N$ is not very efficient.
Remark that, in the grand canonical ensemble, it is then sometimes difficult to pinpoint a precise
value of $\langle N\rangle$ as it requires a fine tuning of the chemical potential $\mu$.

We will concentrate on small number of particles $N$,
which gives a more thorough test of the scaling hypotheses we will introduce
at they should be valid for large $N$.

Using this algorithm, we calculate the average interaction energy $\langle H_{\rm int}\rangle$ 
that gives access to the contact [Eq. (\ref{Hint})]. We choose a system size large enough so density 
becomes zero at the edges of the system.
As the temperature $T$ increases, the simulations become
increasingly difficult: the density distribution of the particles becomes wider,
which means that the events where two particles are superposed and then
contributes to the interaction energy become rare, giving a poor signal to noise
ratio for the contact calculation.
Increasing interactions also reduces the probability of double occupancies and,
consequently, the precision of the calculation.

These difficulties are further enhanced by the fact that,
as $N$ increases, we will
maintain fixed rescaled temperature $\tau$ and interaction $z$ 
to observe possible scaling behaviours. The temperature $T$ and interaction $g$ will then
scale with number of particles as $N$ and $\sqrt{N}$, respectively.
These combined effects strongly limits the temperatures, interactions, and number of 
particles for which we obtain reliable results. For canonical simulations, we were able
to obtain results with a relative error better than two per cent for rescaled interactions up to $z=2.5$, 
rescaled temperatures up to $\tau=5$ and numbers of particles up to $N=5$. Grand canonical
results are more limited. For $N$ up to 4, we are limited to $z=1$ and $\tau=0.2$ if we
want a precision of few percents. For $N=4$, $z=1$ and $\tau=2$, we have relative errors
of order 20\%, which hardly give meaningful information.

%In these cases we then rescale
%the contact by $(N_{\rm target}/\langle N \rangle )^{5/2}$, where $N_{\rm target}$ is 
%the targeted density and where we use the expected
%scaling for the contact in the grand canonical ensemble (see below).
           
\section{Scaling properties}
\label{sec-scaling}
\subsection{Zero temperature scaling}
In \cite{Rizzi2018} we have shown that it is possible
to express the contact for $N$ bosons or $N$ $SU(\kappa)$-fermions
as a function of the contact for two bosons.
Indeed the reduced contact
\begin{equation}
f_N(z,0)=\dfrac{C_N(g(z),0)}{C_N(\infty,0)},
\end{equation}
with $g(z)=2\hbar^2\sqrt{N}z/(ma_{ho})$,
verifies the relation \cite{Rizzi2018}
\begin{equation}
  f_N(z,0)\simeq f_2(z,0), 
\end{equation}
meaning that, upon rescaling of the interaction strength,
all the $N$-dependence of the contact is in $C_N(\infty,0)$.
Moreover it has been shown from a fit on numerical data \cite{Rizzi2018}
that
\begin{equation}
C_N(g(z),0)\sim N^{5/2}-\gamma N^{\eta}
\end{equation}
where $\gamma\simeq 1$ and $\eta=3/4$ in the Tonks-Girardeau limit, 
and where they are slowly varying in the strongly interacting regime $z>1$.

\subsection{Large temperature scaling}
In the large temperature limit, $T\gg T_F$, quantum correlations are
negligible and the contact for $N$ bosons in the canonical ensemble
is simply given by the two-particle contact times the number of pairs
\begin{equation}
  C^c_N(g,T\gg T_F)=\dfrac{N(N-1)}{2}C^c_2(g,T\gg T_F).
  \label{basic}
 \end{equation} 
In the strongly interacting limit Eq. (\ref{basic}) takes the explicit form
(see Appendix)
\begin{equation}
  \begin{split}
C^c_N(z>1,\tau\gg 1)=&\dfrac{N(N-1)}{2}\dfrac{2g}{\pi^{3/2} \hbar\omega a_{ho}^4}\dfrac{1}{\sqrt{\alpha}}\\&
          \left(1-
          \sqrt{\dfrac{\pi}{\alpha}}e^{1/\alpha}
               {\rm Erfc}(1/\sqrt{\alpha})\right)\\
               =&(N^{5/2}-N^{3/2})h_N(z>1,\tau\gg1) \\
  \end{split}
  \label{largecan}
\end{equation}
with $\alpha=4a_{ho}^2\hbar\omega/(\beta g^2)=\tau/z^2$ and
\begin{equation}
h_N(z>1,\tau\gg 1)=\dfrac{2z}{\pi^{3/2} a_{ho}^3}\dfrac{1}{\sqrt{\alpha}}
          \left(1-
          \sqrt{\dfrac{\pi}{\alpha}}e^{1/\alpha}
               {\rm Erfc}(1/\sqrt{\alpha})\right).
               \label{eq:h2inf}
\end{equation}
In the Tonks-Girardeau limit
\begin{equation}
  \begin{split}
    C^c_N(\infty,\tau\gg 1)=&\dfrac{N(N-1)}{2}\dfrac{2}{\pi^{3/2} a_{ho}^3}\sqrt{\dfrac{k_BT}{\hbar\omega}}\\
    =&(N^{5/2}-N^{3/2})h_N(\infty,\tau\gg1)
  \end{split}
\end{equation}
with
\begin{equation}
h_N(\infty,\tau\gg1)=\dfrac{1}{\pi^{3/2} a_{ho}^3}\sqrt{\tau}.
\label{h2inf}
\end{equation}
Analogously to the zero temperature case, we can define the function
\begin{equation}
  f_N(z>1,\tau\gg1)=\dfrac{C_N(g(z),T(\tau))}{C_N(\infty,T(\tau))},
  \label{guess1}
\end{equation}
and we get that
\begin{equation}
  f_N(z>1,\tau\gg1)\simeq f_2(z>1,\tau\gg1)
  \label{guess2}
\end{equation}
holds in the limit $T\gg T_F$,
with $f_2(z>1,\tau\gg 1)=h_2(z>1,\tau\gg 1)/h_2(\infty,\tau\gg1)$.

\subsection{Any temperature scaling conjecture}
\begin{figure}[h]
  \begin{center}
\includegraphics[width=1\linewidth]{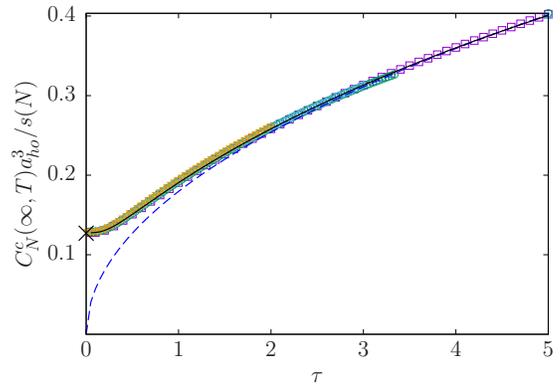}
  \end{center}
  \caption{\label{ganzo}Canonical contact in the Tonks-Girardeau limit $C_N(\infty,T)$,
    Eq. (\ref{miaeq}), as a function of $\tau$,
    scaled by the factor $s(N)=N^{5/2}-N^{3/4(1+\exp(-2/\tau))}$,
    see Eq. (\ref{bellissima}). Violet squares: $N=2$, green circles: $N=3$,
    light-blue up-triangles: $N=4$ and orange down-triangles: $N=5$. The blue
    dashed line corresponds to to the
    high-temperature limit
    $h_2(\infty, \tau \gg 1)$ [Eq. (\ref{h2inf})]. The black cross
  and the black line correspond to $h_2(\infty,0)=(2/\pi)^{3/2}a_{ho}^{-3}(2^{5/2}-2^{3/4})^{-1}$ and $h_2(\infty,\tau)$ [Eq.(\ref{rottura})] respectively.}
\end{figure}

We now propose the general scaling hypothesis that Eq.~(\ref{guess2})
holds for any temperature in the strong-interaction limit.
This is equivalent to claim that, upon rescaling of the interaction strength
and of the temperature, all the $N$-dependence of the contact is embedded
in $C_N(\infty,T)$, for {\it any temperature}.
This dependence is quite trivial at large temperature, as it is
determined by the number of pairs, proportional to $N(N-1)$,
and a $\sqrt{N}$ term that comes from
the rescaling of the temperature with respect to
the Fermi temperature.
By lowering the temperature, the contact almost freezes
at $T\simeq T_F$ and, because of quantum correlations,
there is an enhancement of the dependence on $N$,
from $N^{5/2}-N^{3/2}$ to $N^{5/2}-N^{3/4}$.
This leads us to propose the following conjecture
\begin{eqnarray}
  C_N^c(\infty,\tau)&=&h_2(\infty,\tau)s(N) \label{bellissima}\\
  &=&h_2(\infty,\tau)\left(N^{5/2}-N^{3/4(1+\exp(-2/\tau))}\right),\nonumber
\end{eqnarray}
where
\begin{equation}
  h_2(\infty,\tau)=C_2(\infty,T(\tau))/s(2)
\label{rottura}
\end{equation}
  can de derived by
Eq. (\ref{two-bos-inf}).
In Fig. \ref{ganzo} we plot $C_N^c(\infty,T)$ [Eq. (\ref{miaeq})],
divided by $s(N)$,
as a function of $\tau$, for cases from $N=2$ to $N=5$, as well as
$h_2(\infty, \tau)$, its high-temperature limit $h_2(\infty,\tau \gg 1)$
and its value at zero temperature $h_2(\infty, 0)$.
All the data
collapse on the same curve $h_2(\infty,\tau)$ (continuous black curve),
showing that the
conjecture (\ref{bellissima}) works extremely well.

\begin{figure*}[h]
  \begin{center}
    \includegraphics[width=0.4\linewidth]{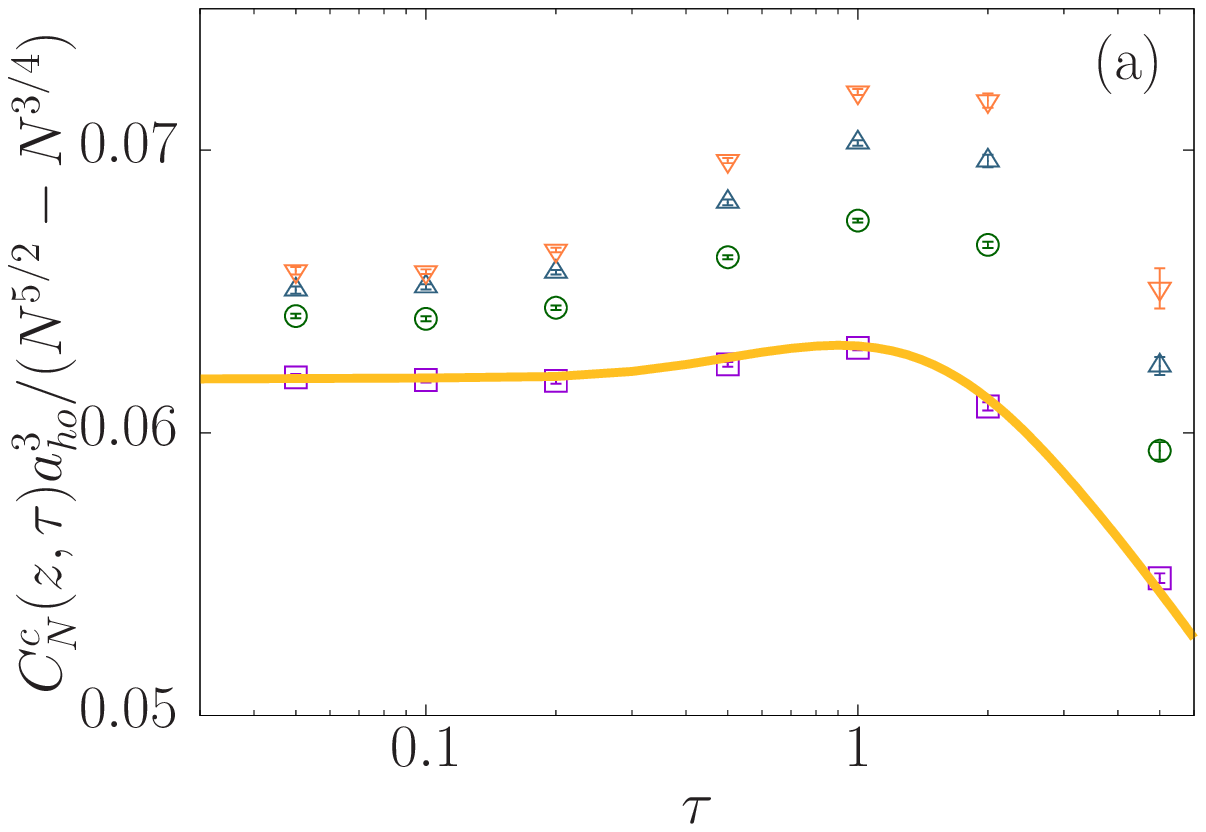}
    \includegraphics[width=0.4\linewidth]{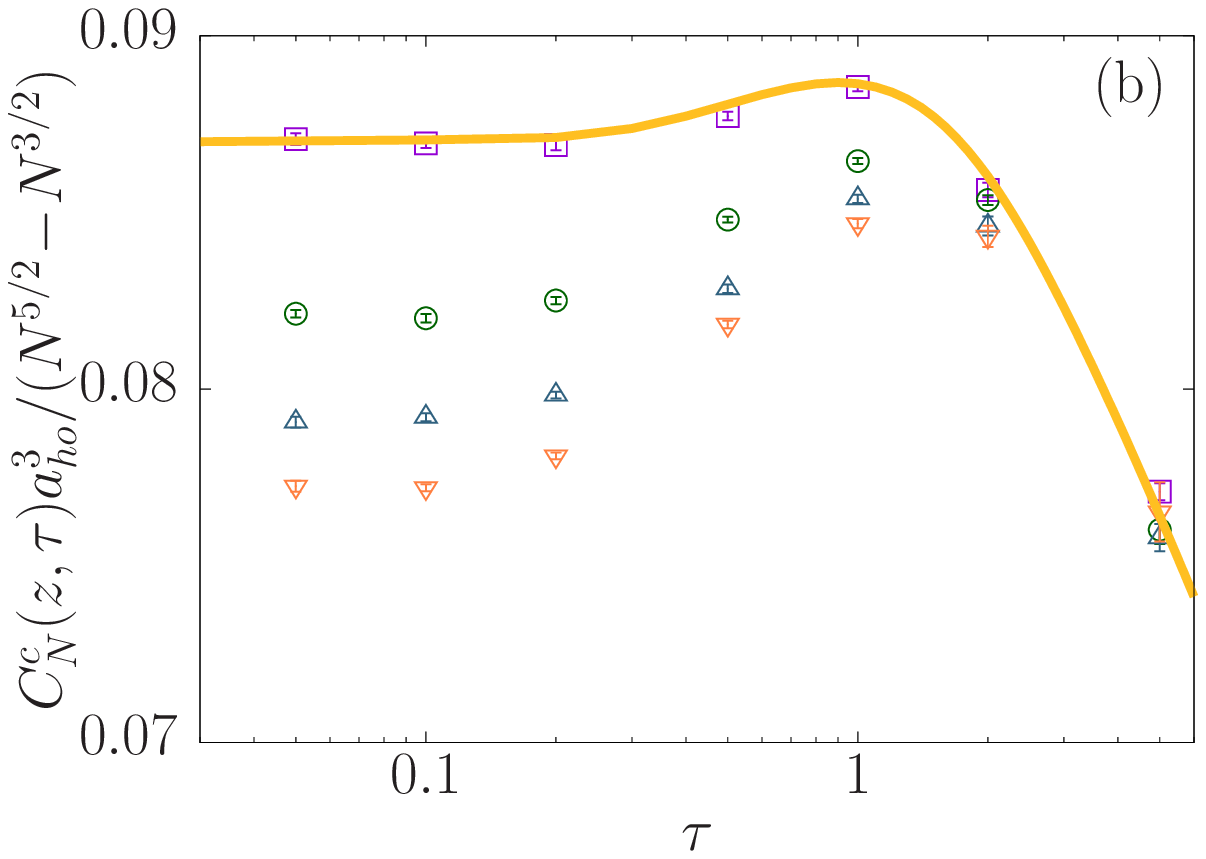}
    \includegraphics[width=0.4\linewidth]{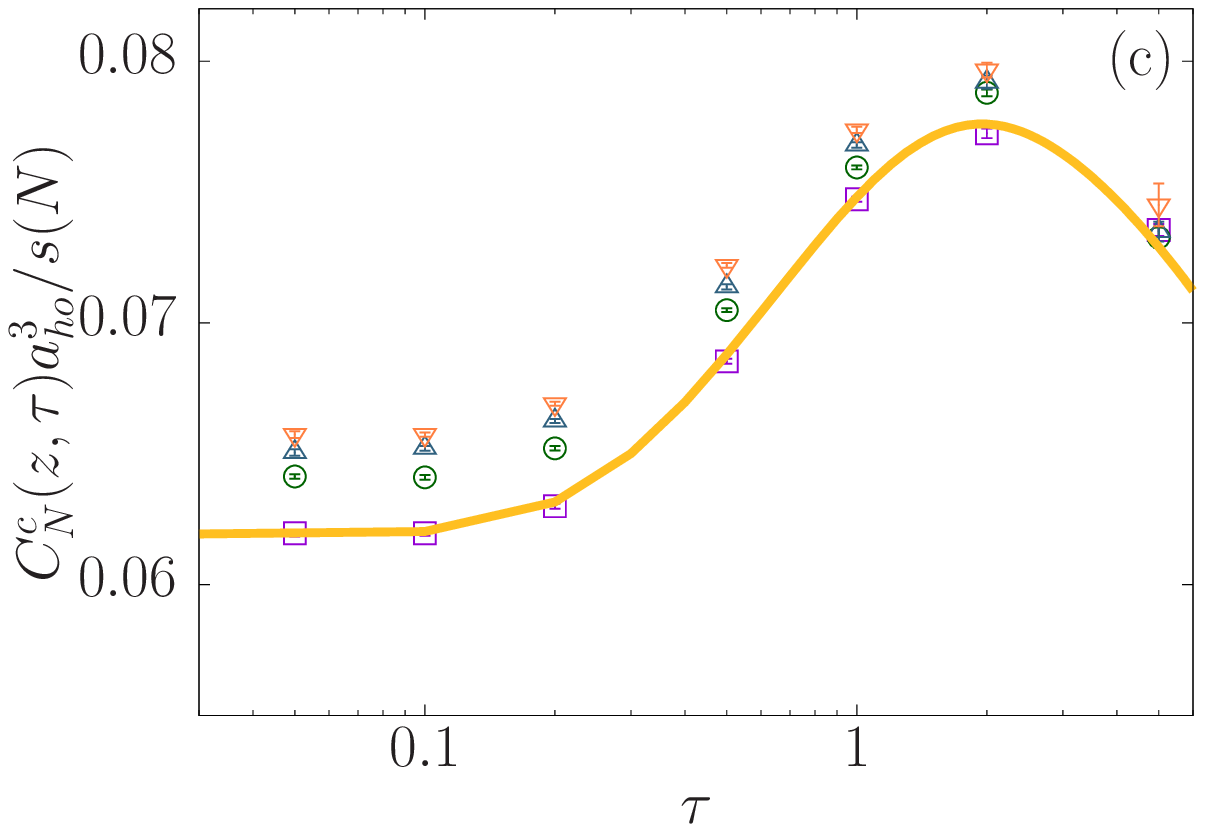}
    \includegraphics[width=0.4\linewidth]{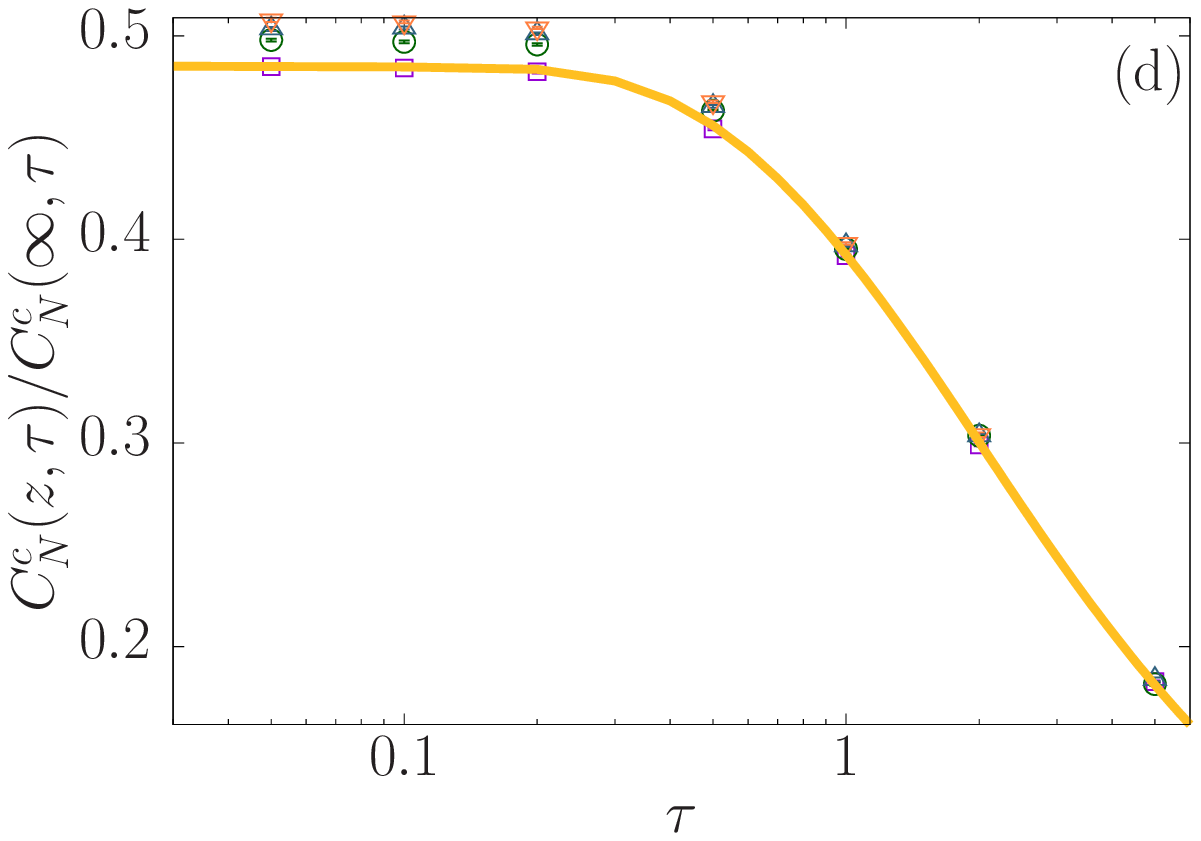}
  \end{center}  
  \caption{\label{can-G2} Panels (a), (b) and (c):
    $C_N^c(z,\tau)a_{ho}^3$ as a function of $\tau$,
    for the case $z=1$, rescaled by $N^{5/2}-N^{3/4}$ (a), $N^{5/2}-N^{3/2}$ (b), and $s(N)$ (c). Panel (d): $f_N(z=1,\tau)$ as a function of $\tau$.
  The points (violet squares: $N=2$, green circles: $N=3$,
  light-blue up-triangles: $N=4$ and orange down-triangles: $N=5$)
  correspond to the QMC data. The continuous yellow line corresponds to the two-bosons
  contact obtained by Eq. (\ref{two-bos}). Non visible QMC error bars are smaller than the symbol size.}
\end{figure*}

\begin{figure*}
   \begin{center}
    \includegraphics[width=0.4\linewidth]{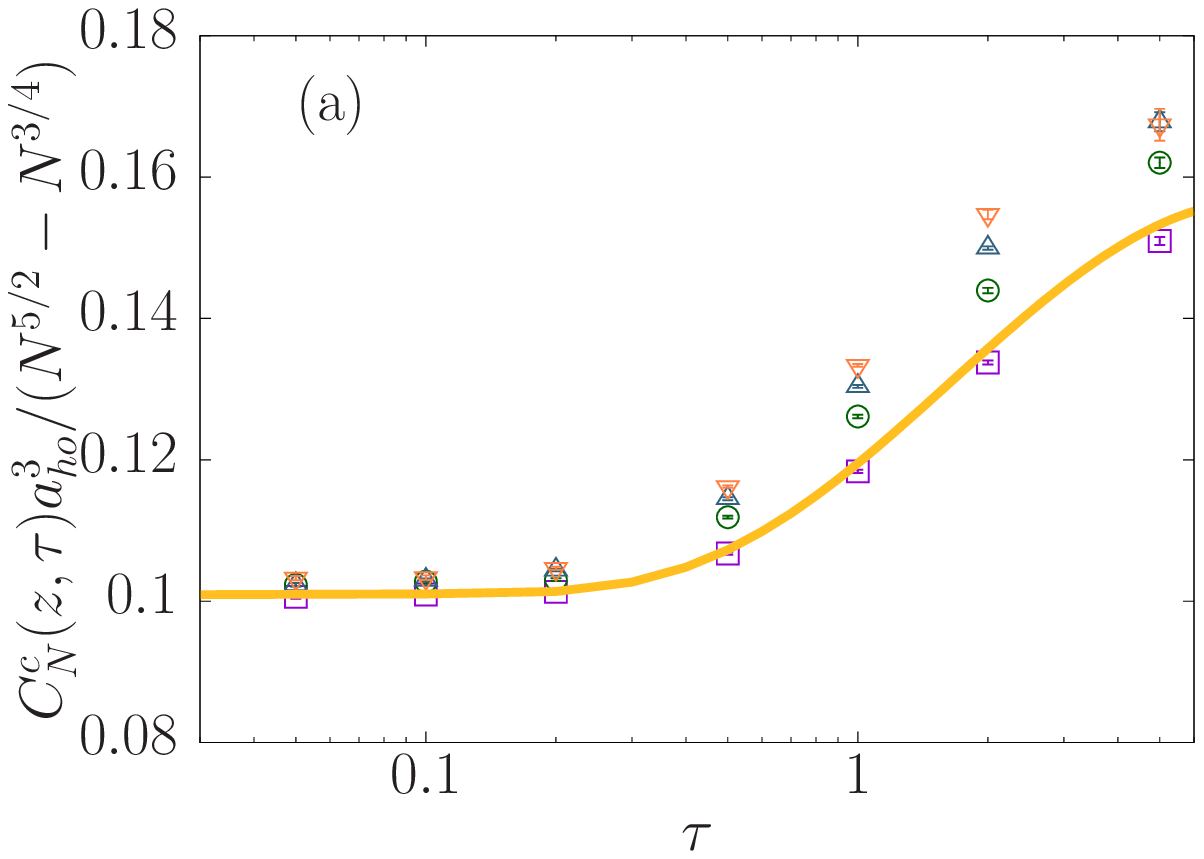}
    \includegraphics[width=0.4\linewidth]{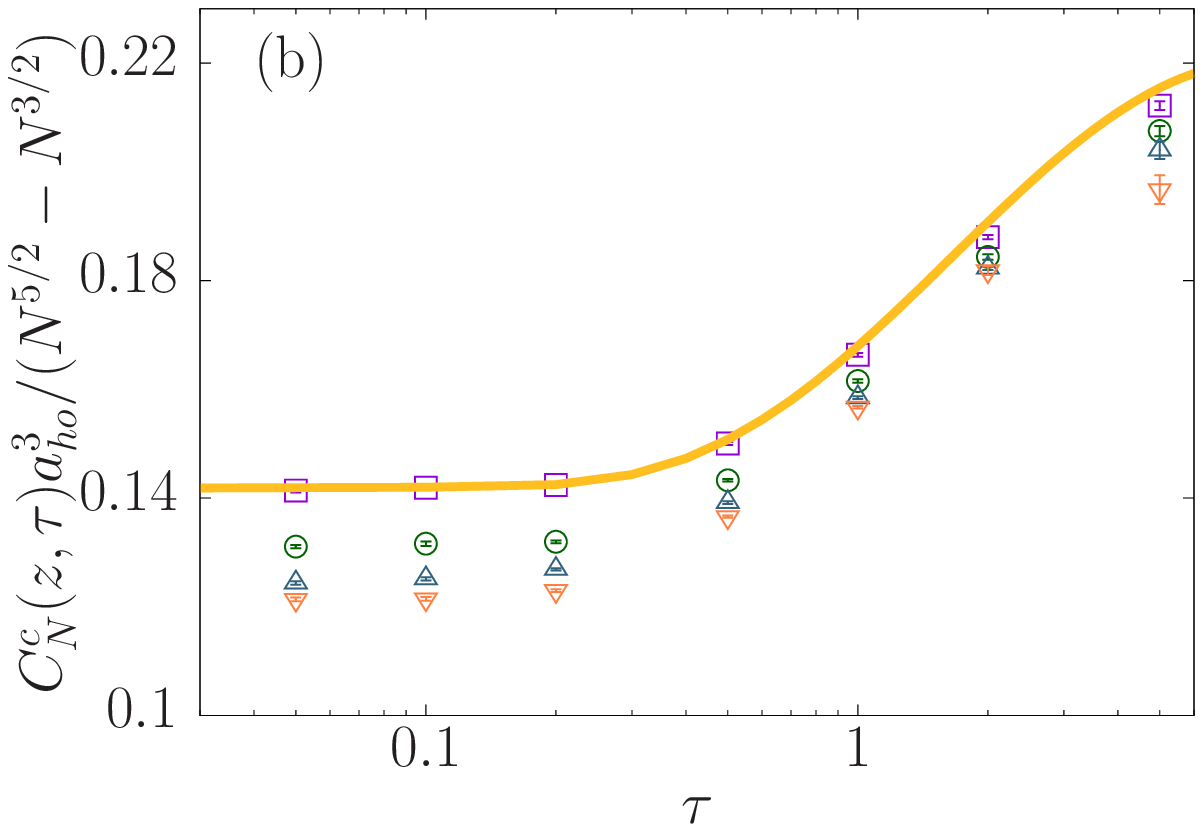}
    \includegraphics[width=0.4\linewidth]{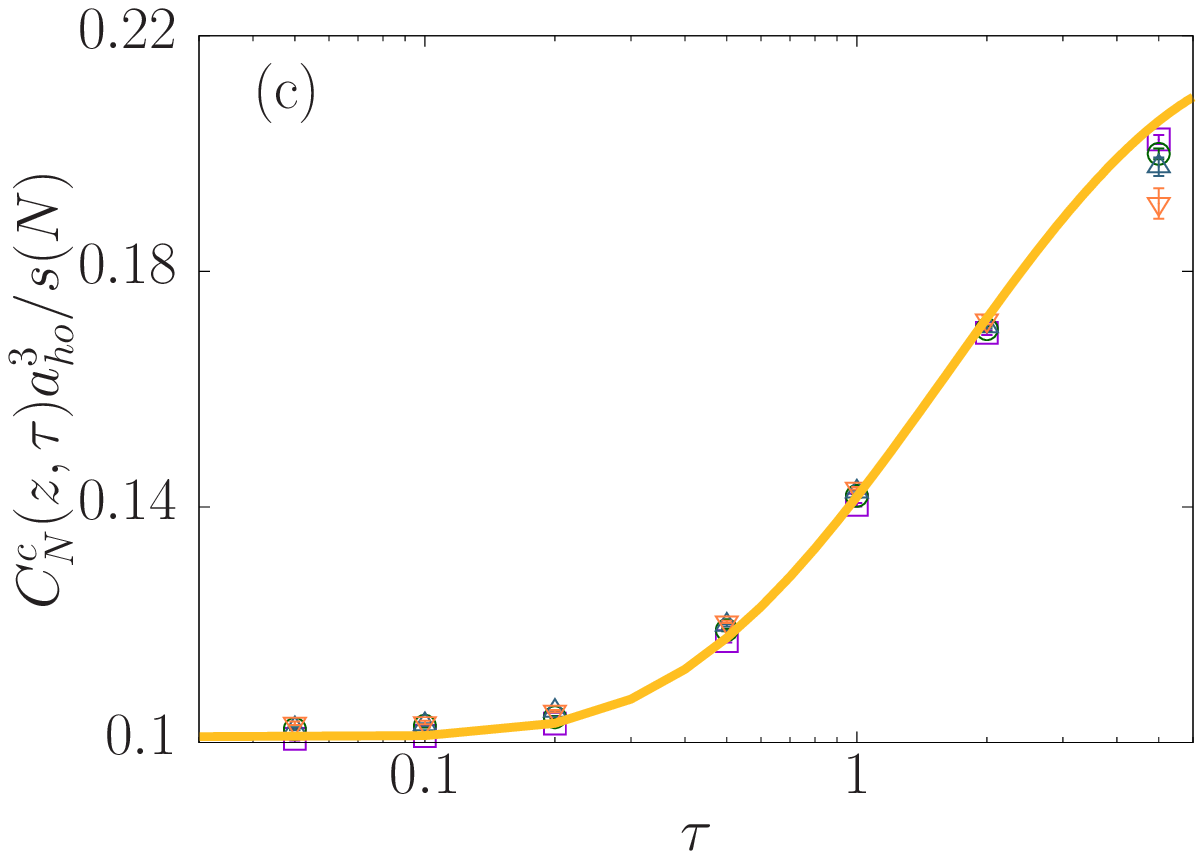}
    \includegraphics[width=0.4\linewidth]{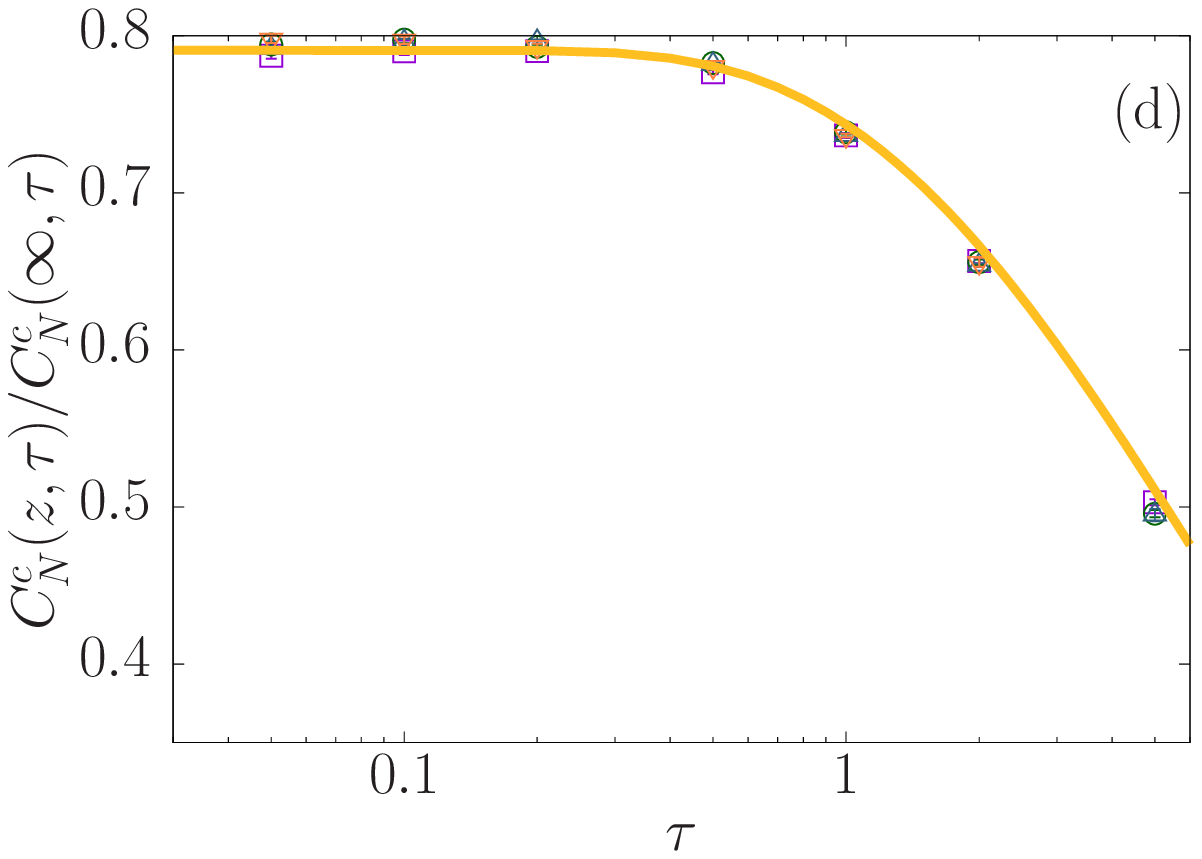}
  \end{center}  
  \caption{\label{can-G5} Panels (a), (b) and (c):
    $C_N^c(z,\tau)a_{ho}^3$ as a function of $\tau$,
    for the case $z=2.5$, rescaled by $N^{5/2}-N^{3/4}$ (a), $N^{5/2}-N^{3/2}$ (b), and $s(N)$ (c). Panel (d): $f_N(z=2.5,\tau)$ as a function of $\tau$.
  The points (violet squares: $N=2$, green circles: $N=3$,
  light-blue up-triangles: $N=4$ and orange down-triangles: $N=5$)
  correspond to the QMC data. The continuous yellow line corresponds to the two-bosons
  contact obtained by Eq. (\ref{two-bos}). Non visible QMC error bars are smaller than the symbol size.}
\end{figure*}

We test now the reliability of the generalized scaling hypothesis
\begin{equation}
  f_N(z>1,\tau)\simeq f_2(z>1,\tau)
  \label{full-scaling}
\end{equation}
approaching the strongly interacting regime.
In Figs. \ref{can-G2} and \ref{can-G5} we plot the canonical contact,
obtained from
quantum Monte-Carlo simulations, for the cases $z=1$
and 2.5, respectively.
For both figures \ref{can-G2} and \ref{can-G5},
in panels (a) the data have been rescaled by $N^{5/2}-N^{3/4}$,
in panels (b) by $N^{5/2}-N^{3/2}$, and in panels (c) by $s(N)$.
The ``zero-temperature'' scaling factor  $N^{5/2}-N^{3/4}$, as obtained in \cite{Rizzi2018}
for the Tonks-Girardeau limit, makes, at small temperatures,
the curves approach at $z=1$ and collapse at $z=2.5$.
The ``pair scaling'' term $N^{5/2}-N^{3/2}$ works well in
the large temperature regime $\tau>1$, while the interpolating function
$s(N)$ [Eq. (\ref{bellissima})] allows the collapse of the data
in the whole temperature range,
with an incertitude of $5\%$ for the case $z=1$ (Fig. \ref{can-G2}-c)
and of $1\%$ for the case $z=2.5$ (Fig. \ref{can-G5}-c).
The validity of the scaling hypothesis (\ref{full-scaling})
is verified in Figs. \ref{can-G2}-d and \ref{can-G5}-d.
Remark that, as mentioned earlier, precise QMC results are
limited to small number of particles and intermediate values of $\tau$ and $z$. 
The limitation on the number of particles is not crucial as,
for large number of particles,
$\lim_{N\rightarrow\infty}s(N)/N^{5/2}=1$, and we recover
the known thermodynamics limit. Concentrating on small number
of particles $N\le5$ then provides a more stringent verification  of
the reliability of the scaling hypothesis (\ref{full-scaling}).

\section{Comparison with the grand-canonical Tan's contact}
\label{sec-comp}
In the zero temperature limit, the grand-canonical and canonical
contacts coincide, thus, in the strongly
interacting regime, both scale as $\sim (N^{5/2}-N^{3/4})$.

But, as soon as the temperature increases, the grand-canonical contact
for an average number $\langle N\rangle$ of particles 
departs from the canonical one for $N$ particles.
Indeed, with larger numbers contributions, the grand-canonical
contact increases more rapidly than the canonical
one that is almost constant
for $0\le\tau\le0.5$ (see Fig. \ref{Tonks-c-gc} for the Tonks-Girardeau limit case).

\begin{figure}
\begin{center}
  \includegraphics[width=1\linewidth]{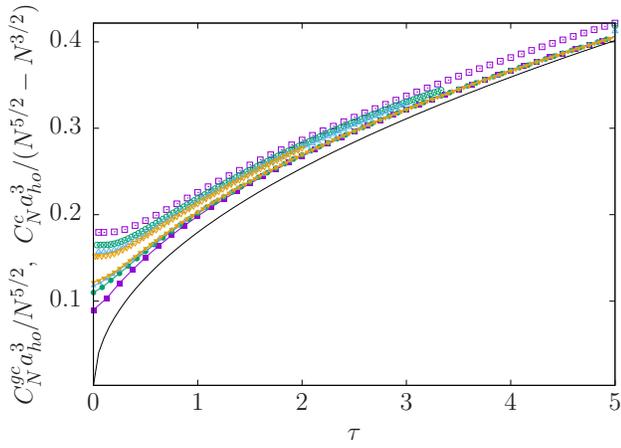}
  \caption{\label{fig:superscaling} Canonical (empty symbols)
    and grand-canonical contact (full symbols) as a function of $\tau$
    for $N=2$ (violet squares),
    $N=3$ (green circles), $N=4$ (light-blue up-triangles), and $N=5$ (orange down-triangles) Tonks-Girardeau bosons.
   The canonical contact [Eq.~(\ref{miaeq})] is rescaled by a factor
   $N^{5/2}-N^{3/2}$, while the grand-canonical one
   [Eqs. (8)-(9) in \cite{vignolo2013}] is rescaled by $N^{5/2}$.
  The black continuous curve corresponds to $\sqrt{\tau}/\pi^{3/2}$ [Eq. (\ref{h2inf})].}
    \end{center}
      \end{figure}

\begin{figure}
\begin{center}
  \includegraphics[width=1\linewidth]{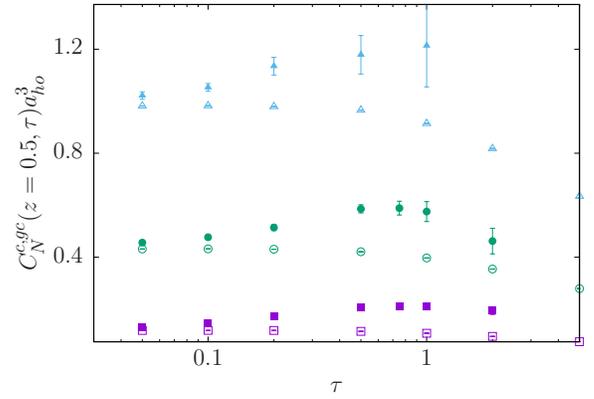}
  \caption{\label{dis} Canonical (empty symbols)
    and grand-canonical contact (full symbols) as a function of $\tau$
    for $N=2$ (violet squares),
    $N=3$ (green circles), $N=4$ (light-blue up-triangles) bosons.
    All points correspond to QMC data evaluated in
    the weakly-interacting regime $z=0.5$. QMC error bars for the canonical data are smaller than the symbol size.}
    \end{center}
      \end{figure}

    In the large temperature limit, in the grand-canonical ensemble,
    the term  $N(N-1)$, proportional to the number of pairs in the
    canonical ensemble,
    has to be replaced by its average value
\begin{equation}
    \langle N(N-1)\rangle=\langle N^2\rangle-\langle N\rangle=
    \langle N\rangle^2.
\end{equation}
This follows from the fact that, at large $T$,
$\langle \Delta N^2\rangle\simeq \langle N\rangle$.
By defining $T_F=\langle N\rangle\hbar\omega/k_B$,
we find
\begin{equation}
C_N^{gc}(g,T\gg T_F)=\dfrac{\langle N\rangle ^2}{2}C_2^c=\langle N\rangle ^{5/2}
      h_2(z>1,\tau\gg 1),
\end{equation}
in agreement with the virial calculation \cite{Yao2018}.
Thus, in the large temperature limit,
$C_N^{gc}(g,T\gg T_F)/\langle N\rangle ^{5/2}$ and
$C_N^{c}(g,T\gg T_F)/(N^{5/2}-N^{3/2})$ collapse on the same curve
$h_2(z,\tau\gg 1)=\sqrt{\tau}/(\pi^{3/2}a_{ho}^3)$.
This is shown in Fig.~\ref{fig:superscaling} for the Tonks-Girardeau limit,
where we have compared the canonical contact [Eq.~(\ref{miaeq})]
and the grand-canonical one as obtained from
Eqs. (8)-(9) in \cite{vignolo2013}.
Remark that the convergence is faster for the grand-canonical contact.
    The consequence of the fact that the canonical and the grand-canonical
    contact are proportional to one another, at large temperature $\tau\gg1$,
    is that
    both have a maximum at
    $\tau=1.48 z^2$ in the strong-interacting limit \cite{Yao2018}.
    The situation is different in the weak-interaction regime, where the
    grand-canonical contact exhibits a maximum at lower temperatures.
    This maximum, that has been explained as the mark of the crossover
    between a quasi-condensate and an ideal Bose gas \cite{Yao2018},
    is not present in the canonical case.
    This has been studied by means of QMC simulations and
    shown in Fig. \ref{dis}.

In the canonical ensemble and at low interactions the contact decreases with
increasing temperature because, as particles occupy individual excited states,
the cloud of particles spreads and the interaction energy is lowered.
This happens when the temperature is large enough to overcome
the $\hbar \omega$ gap between the ground and excited states, which
explains why there is almost no variation at low temperature.

In the grand canonical ensemble, the same effect will of course take
place and yields to the same decrease of the contact at high temperature.
However, at low temperature, another phenomena occurs: the probability 
to have a number of particles that is larger than $\langle N\rangle$ increases
with temperature. This gives larger contributions to the interaction energy
and explains the initial increase of the contact at low temperatures.

\begin{figure}
  \begin{center}
    \includegraphics[width=1\linewidth]{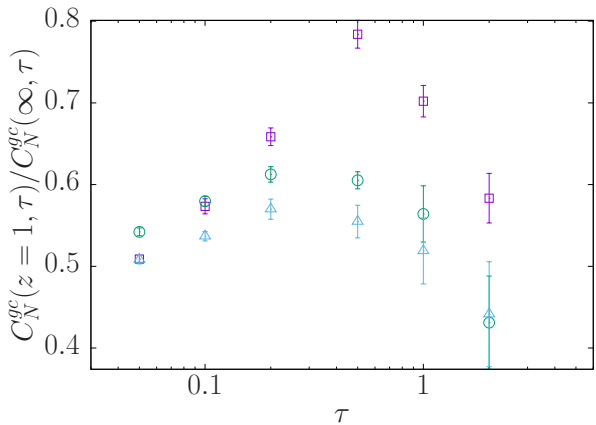}
  \end{center}
  \caption{\label{gc-scaling}$C_N^{gc}(z=1,\tau)/C_N^{gc}(\infty,\tau)$
    as a function of $\tau$.
  The points (violet squares: $N=2$, green circles: $N=3$,
  light-blue up-triangles: $N=4$)
  correspond to the QMC data.}
  \end{figure}

As Eq. (\ref{guess2}) holds even in the grand-canonical ensemble,
one may wonder
if the generalized scaling hypothesis (\ref{full-scaling}) is still
valid in this ensemble.
In Fig. \ref{gc-scaling} we plot the quantity
$C_N^{gc}(z,\tau)/C_N^{gc}(\infty,\tau)$
for the case $z=1$ and $N=2$, 3 and 4 and $\tau\le 2$,
$C_N^{gc}(z,\tau)$ having been calculated by means of QMC simulations and
$C_N^{gc}(\infty,\tau)$ by means of Eqs. (8)-(9) in \cite{vignolo2013}.
We observe that, for small and intermediate temperatures, in the intermediate interactions regime,
the curves remain different, 
instead of the collapse observed in the canonical case (see Fig. \ref{can-G2}(d)).
Our scaling hypothesis then fails in this case of intermediate interactions,
as the grand-canonical Tonks-Girardeau contact does not embed
the full $\langle N\rangle$-dependency for these intermediate
interactions. We were not able to test this scaling hypothesis in the grand canonical 
ensemble at larger interactions as QMC simulations become increasingly difficult.
\section{Conclusion}
\label{sec-concl}
In this paper we have shown that
the canonical contact for $N$, harmonically trapped, Lieb-Liniger bosons,
at any temperature, in the
  repulsive strongly interacting regime, can be written as a function of the
  two-bosons contact and the contact for $N$ Tonks-Girardeau bosons.
  The first can be easily calculated and we provide an analytical formula
  for the second for any number of bosons and temperature. This enlightens
  the dependence of the contact on the number of pairs at large temperature
  and the effects of
  correlations at low temperature. Moreover, it supplies a scaling function,
  in the canonical ensemble, 
  for any number of particles $N\geq2$ and any temperature in the strong interacting regime.
  We have proven our theory for small number of bosons ($2\le N\le 5$) where
  corrections with respect to the known thermodynamic limit are more important.
  We have been informed that these results may also hold true
  for a 1D homogeneous Bose gas. This can be deduced from
  the results recently presented in \cite{DeRosi2019}.
  In this paper the authors show that in the strongly interacting limit
  $C_N^c = 4 m N P_H/\hbar^2$.
   The force $P_H$ is expressed as $P_H=n^3f_H(z_H,\tau_H)$, where
   $z_H=(n a_{1D})^{-1}$ is the
   rescaled interaction strength for the homogeneous system of linear
   density $n$, $\tau_H=T/T_{F,H}$ is the rescaled temperature
   ($T_{F,H}$ being the Fermi temperature for the homogeneous system),
   and $f_H$ is a universal function of  $z_H$ and  $\tau_H$.
   From this it can be deduced that $C_N^c(z_H>1,\tau_H)/C_N^c(\infty,\tau_H)$
   is also a universal function, which is equivalent for an homogeneous
   system of the scaling relations found in the trapped case.
     
  Finally we discuss the difference between the canonical and grand-canonical
  contacts. At large temperature these quantities are both proportional to
  the two bosons contact, and the proportionality factor
  depends on the number of pairs
  in the canonical ensemble and the average number of pairs in the
  grand-canonical one. The main difference between the grand-canonical and
  canonical cases is that, at small and intermediate temperatures,
  the grand-canonical contact for $\langle N\rangle$ bosons
   cannot be written as a function of the
 $\langle 2\rangle$-bosons contact and the contact for $\langle N\rangle$
   Tonks-Girardeau bosons, as far as we can test it with the QMC simulations
   in the intermediate interaction regime. 
   Namely, at variance from the canonical case,
   the  grand-canonical contact for $\langle N\rangle$
   Tonks-Girardeau bosons
  seems not to embed the dependence for the average number
  of particles $\langle N\rangle$. Indeed our scaling hypothesis fails
  as far as we can test it with the QMC simulations in the intermediate
  interaction regime.
  
  Our work can be relevant for experiments with a small number of
  particles \cite{Zurn2012,Wenz2013}.
  From a conceptual point of view, it is an important step forward in
  understanding the effects of correlations and interactions in
  finite-temperature harmonically trapped one-dimensional bosons,
  as well as in enlightening the role of the particle-number fluctuations.
  The extension to the case of multi-component systems is not
  straightforward and will be the subject of a further study.

  \begin{acknowledgments}
    We thank an anonymous referee for many useful suggestions and, in
    particular, for making us aware that our results hold even in the 1D untrapped Bose gas.
P.V. thanks A. Minguzzi for useful discussions.
The work of F.T.S. was financed in part by the Coordena\c{c}\~ao de Aperfei\c{c}oamento
de Pessoal de N\'{\i}vel Superior - Brasil (CAPES) - Finance Code 001.
\end{acknowledgments}

\bigskip
 
    \appendix

    \section{Two-body contact in the strong
      interaction and large temperature limit}
      We start with Eq. (\ref{two-bos})
    \begin{equation}\label{app1}
	    C_2^{c}=-\dfrac{m^2\omega}{\pi\hbar^3} Z_r^{-1} \sum_n e^{-\beta\hbar\omega\nu_n}\dfrac{\partial\nu_n}{\partial g^{-1}}.
      \end{equation}
    It can be shown \cite{Yao2018} that, in the strongly interacting limit
    $z>1$,
      the solutions of Eq. (\ref{eq.gamma}) are given by
      \begin{equation}
        \nu_n\simeq\dfrac{2}{\pi}{\rm acot}(2\sqrt{2n+1}g^{-1}\hbar\omega a_{ho})+2n,
        \label{pioggia}
      \end{equation}
      with $n\ge 0$. This approximation (\ref{pioggia}) becomes more precise at large
      values of $n$.
      Thus (\ref{app1}) reads
      \begin{equation}
      C_2^{c}=\dfrac{4 Z_r^{-1}}{\pi^2 a_{ho}^3} \sum_n \dfrac{e^{-\beta\hbar\omega\nu_n} \sqrt{2n+1}}{1+4(2n+1)(\hbar\omega a_{ho}g^{-1})^{2}}.\end{equation}
        
      By replacing in the exponential $\nu_n$ with its value
      in the Tonks-Girardeau limit, $\nu_n=2n+1$, and exploiting that
        \begin{equation}
          \begin{split}
		  &\int_0^\infty\dfrac{\sqrt{x}}{1+xb^2}e^{-\beta\hbar\omega x}{\rm d}x\\
		  &=
            \dfrac{1}{(\beta\hbar\omega)^{3/2}}\dfrac{\sqrt{\pi}}{\alpha}\left(1-
            \sqrt{\dfrac{\pi}{\alpha}}e^{1/\alpha}
            {\rm Erfc}(1/\sqrt{\alpha})\right),\\
\end{split}
          \end{equation}
        with $\alpha=b^2/(\hbar\omega\beta)=4a_{ho}^2\hbar\omega/(\beta g^2)$, we have that
        \begin{equation}
          C_2^{c}=\dfrac{2g}{\pi^{3/2} \hbar\omega a_{ho}^4}\dfrac{1}{\sqrt{\alpha}}
          \left(1-
            \sqrt{\dfrac{\pi}{\alpha}}e^{1/\alpha}{\rm Erfc}(1/\sqrt{\alpha})\right).\label{diluvio}
        \end{equation}
        Remark that Eq. (\ref{diluvio}) is valid only in the large temperature
        limit where replacing the sum with an integral is a valid approximation.
        Hence, in the Tonks-Girardeau limit, the contact reduces to
        \begin{equation}\lim_{g\rightarrow\infty}C_2^c=
          \dfrac{2}{\pi^{3/2} a_{ho}^3}\sqrt{\dfrac{k_BT}{\hbar\omega}}.
        \end{equation}

%%%%%%%%%%%%%%%%%%%%%%%%%%%%%5        
%        \bibliographystyle{prsty}
%        \bibliography{biblioferm}

\begin{thebibliography}{10}

\bibitem{Lieb1963}
E. Lieb and W. Liniger, Phys. Rev. {\bf 130},  1605  (1963).

\bibitem{McGuire1964}
J.~B. McGuire, J. Math. Phys. (NY) {\bf 5},  622  (1964).

\bibitem{Yang1969}
C. Yang and C. Yang, J. Math. Phys. {\bf 130},  1605  (1969).

\bibitem{LiebLiniger}
E. Lieb, Phys. Rev. {\bf 130},  1616  (1963).

\bibitem{Yang67}
C.~N. Yang, Phys. Rev. Lett. {\bf 19},  1312  (1967).

\bibitem{Gaudin1967}
M. Gaudin, Physics Letters A {\bf 24},  55   (1967).

\bibitem{Sutherland68}
B. Sutherland, Phys. Rev. Lett. {\bf 20},  98  (1968).

\bibitem{LaiYang}
C.~K. Lai and C.~N. Yang, Phys. Rev. A {\bf 3},  393  (1971).

\bibitem{McGuire}
J.~B. McGuire, J. Math. Phys. (NY) {\bf 6},  432  (1965).

\bibitem{Luther1974}
A. Luther and V.~J. Emery, Phys. Rev. Lett. {\bf 33},  589  (1974).

\bibitem{Fuchs2004}
J.~N. Fuchs, A. Recati, and W. Zwerger, Phys. Rev. Lett. {\bf 93},  090408
  (2004).

\bibitem{Busch98}
T. Busch, B.-G. Englert, K. Rz\c{a}\.{z}ewski, and M. Wilkens, Found. Phys.
  {\bf 28},  549  (1998).

\bibitem{Aharony2000}
A. Aharony, O. Entin-Wohlman, and Y. Imry, Phys. Rev. B {\bf 61},  5452
  (2000).

\bibitem{Abad2005}
J. Abad and J.~G. Esteve, Few-Body Systems {\bf 37},  107  (2005).

\bibitem{Olshanii03}
M. Olshanii and V. Dunjko, Phys. Rev. Lett. {\bf 91},  090401  (2003).

\bibitem{Yao2018}
H. Yao {\it et~al.}, Phys. Rev. Lett. {\bf 121},  220402  (2018).

\bibitem{Giamarchi_book}
T. Giamarchi, {\em Quantum Physics in One Dimension} (Clarendon Press, Oxford,
  2003).

\bibitem{Gross1961}
E. Gross, Il Nuovo Cimento {\bf 20},  454  (1961).

\bibitem{Pitaevskii1961}
L. Pitaevskii, Sov. Phys. JEPT {\bf 13},  451  (1961).

\bibitem{Tan2008a}
S. Tan, Ann. Phys. (N.Y.) {\bf 323},  2971  (2008).

\bibitem{Tan2008b}
S. Tan, Ann. Phys. (N.Y.) {\bf 323},  2987  (2008).

\bibitem{Tan2008c}
S. Tan, Ann. Phys. (N.Y.) {\bf 323},  2952  (2008).

\bibitem{Stewart2010}
J.~T. Stewart, J.~P. Gaebler, T.~E. Drake, and D.~S. Jin, Phys. Rev. Lett. {\bf
  104},  235301  (2010).

\bibitem{Sagi2012}
Y. Sagi, T.~E. Drake, R. Paudel, and D.~S. Jin, Phys. Rev. Lett. {\bf 109},
  220402  (2012).

\bibitem{Chang2016}
R. Chang {\it et~al.}, Phys. Rev. Lett. {\bf 117},  235303  (2016).

\bibitem{Wild2012}
R.~J. Wild {\it et~al.}, Phys. Rev. Lett. {\bf 108},  145305  (2012).

\bibitem{Yan2019}
Z. Yan {\it et~al.}, Phys. Rev. Lett. {\bf 122},  093401  (2019).

\bibitem{Hoinka2013}
S. Hoinka {\it et~al.}, Phys. Rev. Lett. {\bf 110},  055305  (2013).

\bibitem{Laurent2017}
S. Laurent {\it et~al.}, Phys. Rev. Lett. {\bf 118},  103403  (2017).

\bibitem{Decamp2016b}
J. Decamp {\it et~al.}, Physical Review A {\bf 94},  053614  (2016).

\bibitem{Decamp2017}
J. Decamp {\it et~al.}, New Journal of Physics {\bf 19},  125001  (2017).

\bibitem{Minguzzi02}
A. Minguzzi, P. Vignolo, and M. Tosi, Phys. Lett. A {\bf 294},  222  (2002).

\bibitem{Lewenstein-Massignan}
T. Grining {\it et~al.}, Phys. Rev. A {\bf 92},  061601  (2015).

\bibitem{Matveeva2016}
N. Matveeva and G. Astrakharchik, New Journal of Physics {\bf 18},  065009
  (2016).

\bibitem{Patu2016}
O.~I. P\^a\ifmmode~\mbox{\c{t}}\else \c{t}\fi{}u and A. Kl\"umper, Phys. Rev. A
  {\bf 93},  033616  (2016).

\bibitem{xu2015}
W. Xu and M. Rigol, Phys. Rev. A {\bf 92},  063623  (2015).

\bibitem{Rizzi2018}
M. Rizzi, C. Miniatura, A. Minguzzi, and P. Vignolo, Phys. Rev. A {\bf 98},
  043607  (2018).

\bibitem{Moritz2003}
H. Moritz, T. St\"oferle, M. K\"ohl, and T. Esslinger, Phys. Rev. Lett. {\bf
  91},  250402  (2003).

\bibitem{Pagano2014}
G. Pagano {\it et~al.}, Nature Physics {\bf 10},  198–201  (2014).

\bibitem{Salces2018}
F. Salces-Carcoba {\it et~al.}, New Journal of Physics {\bf 20},  113032
  (2018).

\bibitem{Olsh98}
M. Olshanii, Phys. Rev. Lett. {\bf 81},  938  (1998).

\bibitem{Valiente2012}
M. Valiente, European Phys. Lett. {\bf 98},  10010  (2012).

\bibitem{Gradshteyn}
I. Gradshteyn and I. Ryzhik, {\em Table of Integrals, Series, and Products}
  (Elsevier, Amsterdam, 1996).

\bibitem{Pitaevskii2016}
L. Pitaevskii and S. Stringari, {\em Bose-Einstein Condensation and
  Superfluidity} (Oxford University Press, Oxford, 2016).

\bibitem{Fang09}
B.~Y. Fang, P. Vignolo, C. Miniatura, and A. Minguzzi, Phys. Rev. A {\bf 79},
  023623  (2009).

\bibitem{vignolo2013}
P. Vignolo and A. Minguzzi, Phys. Rev. Lett. {\bf 110},  020403  (2013).

\bibitem{rousseau08}
V.~G. Rousseau, Phys. Rev. E {\bf 77},  056705  (2008).

\bibitem{rousseau08b}
V.~G. Rousseau, Phys. Rev. E {\bf 78},  056707  (2008).

\bibitem{DeRosi2019}
G.~D. Rosi, P. Massignan, M. Lewenstein, and G. Astrakharchik, arXiv:1905.07391
   (2019).

\bibitem{Zurn2012}
G. Z\"urn {\it et~al.}, Phys. Rev. Lett. {\bf 108},  075303  (2012).

\bibitem{Wenz2013}
A.~N. Wenz {\it et~al.}, Science {\bf 342},  457  (2013).

\end{thebibliography}

\end{document}